\documentclass[aps,prapplied,reprint,superscriptaddress,longbibliography]{revtex4-2}

\usepackage{amsmath,amssymb,bm}
\usepackage{graphicx}
\usepackage{physics}
\usepackage{mathtools}
\usepackage{xcolor}
\usepackage{textcomp}   
\usepackage{nameref}
\usepackage{hyperref}
\usepackage{booktabs}   
\usepackage{tabularx}
\usepackage{array}
\usepackage{xcolor}

\newcommand{\SCNU}{\affiliation{1}{Key Laboratory of Atomic and Subatomic Structure and Quantum Control (South China Normal University), Ministry of Education, Guangdong Basic Research Center of Excellence for Structure and Fundamental Interactions of Matter, School of Physics, South China Normal University, Guangzhou 510006, China}}

\newcommand{\IQA}{\affiliation{2}{International Quantum Academy, Shenzhen 518048, China}}
\newcommand{\SZU}{\affiliation{3}{Institute of Micro and Nano Optoelectronics, Shenzhen University, Shenzhen 518060, China}}

\newcommand{\SUSTECH}{\affiliation{4}{Southern University of Science and Technology, Shenzhen 518055, China}}
\newcommand{\USTC}{\affiliation{5}{School of Emerging Technology, University of Science and Technology of China, Anhui 230026, China}}
\newcommand{\HFNL}{\affiliation{6}{International Quantum Academy, and Shenzhen Branch, Hefei National Laboratory, Shenzhen 518048, China}}

\begin{document}

\title{Broadband Purcell Filter for Fast Superconducting Qubit Reset and Readout}

\author{Yu Zhao}
\affiliation{\SCNU}\affiliation{\IQA}

\author{Zhixu Chen}
\affiliation{\IQA}\affiliation{\USTC}
\author{Hanxian Liu}
\affiliation{\IQA}\affiliation{\SUSTECH}
\author{Mingze Liu}
\affiliation{\IQA}\affiliation{\SUSTECH}
\author{Zixing Liu}
\affiliation{\IQA}\affiliation{\SZU}
\author{Hao Pang}
\affiliation{\IQA}\affiliation{\SUSTECH}
\author{Meiyan Wan}
\affiliation{\IQA}
\author{Changkun Wu}
\affiliation{\IQA}\affiliation{\SUSTECH}
\author{Liuzhu Zhong}
\affiliation{\IQA}\affiliation{\SUSTECH}
\author{Sai Li}
\affiliation{\SCNU}

\author{Yuefeng Yuan }
\affiliation{\IQA}

\author{Yuxuan Zhou}
\email{zhouyuxuan@iqasz.cn}
\affiliation{\IQA}

\author{Ji Jiang}
\email{jiangji@iqasz.cn}
\affiliation{\IQA}

\author{Ji Chu}
\email{jichu@iqasz.cn}
\affiliation{\IQA}

\author{Song Liu}
\affiliation{\IQA}\affiliation{\HFNL}

\date{\today}

\begin{abstract}
Rapid reset and readout of qubit states are essential for quantum error correction, yet accelerating these operations through stronger coupling to a dissipative environment inevitably increases qubit decay via the Purcell effect. Here we present a broadband Purcell filter that decouples the reset and readout paths, enabling both operations to be independently optimized without compromising qubit coherence. The filter employs two engineered notches—an intrinsic notch and a bandstop notch—to provide broadband Purcell protection, together with an additional reset stub that creates a reset mode below the protected band. 
To enable fast reset while suppressing filter-mediated interactions between qubits, we couple each qubit to a dedicated reset resonator.
We experimentally demonstrate Purcell-limited relaxation times exceeding 1~ms across a 1.2~GHz bandwidth, simultaneously with 500~ns readout without a Josephson parametric amplifier and 100~ns reset with 99.6\% efficiency. The reset resonator is designed with a deliberate $\kappa$–$\chi$ mismatch, which suppresses photon-shot-noise-induced dephasing by a factor of 70 compared to the readout resonator. Our work provides a scalable hardware solution that resolves the traditional trade-off between fast qubit operations and qubit protection, advancing the prospects for fault-tolerant quantum computing.
\end{abstract}

\maketitle

\section{Introduction}

Quantum error correction (QEC) is a key enabling technology for scalable, fault-tolerant quantum computing~\cite{knill1997theory,kitaev2003fault,terhal2015quantum}. Significant experimental progress has been made in recent years using superconducting qubits~\cite{google2023suppressing,google2025quantum,he2025experimental}. 
A prerequisite for scalable QEC is the ability to rapidly reset and read out qubit states. Accelerating these operations requires strong coupling to a dissipative environment, which increases the qubit decay rate through the resonator into the output line. 
Purcell filters address this trade-off by shaping the circuit impedance to allow strong coupling to the
environment at the readout frequency while suppressing coupling at the qubit frequency~\cite{reed2010fast,sete2014purcell}.
Shared Purcell filters~\cite{jeffrey2014fast}, including their high-order variants~\cite{bronn2015broadband,sunada2022fast,smitham2025sub,zhou2024high,yan2023broadband,park2024characterization,bakr2025intrinsic,spring2025fast,escribano2026engineered,luo2026compact}, are widely used for rapid readout and broadband $T_1$ protection.

For qubit reset, however, the conventional approach uses the same readout resonator as the dissipation channel, which imposes two constraints:
(1) For multi-level reset, one must choose either a resonator-below-qubit configuration~\cite{mcewen2021removing}, which is vulnerable to measurement-induced state transitions~\cite{Sank2016,khezri2023measurement}, or a resonator-above-qubit configuration, which requires either complicated control~\cite{egger2018pulsed,magnard2018fast,zhou2021rapid,marques2023all,Yang2024_PRL,lacroix2025fast,battistel2021hardware} or additional hardware~\cite{he2025experimental}.
(2) The parameters optimal for high-fidelity readout are incompatible with those for reset: while $\kappa$-$\chi$ matching maximizes the signal-to-noise ratio (SNR), the residual photons remaining in the resonator after a qubit-resonator iSWAP operation induce significant qubit dephasing, limiting the reset speed because the system must wait for the resonator photons to deplete.
Multi-purpose architectures decoupled readout and reset have been proposed~\cite{ding2025multipurpose,gu2026multimode}, but their practicality is limited by either structural complexity or stringent parameter constraints.

\begin{figure*}[t]
    \centering
    \includegraphics[width=\textwidth]{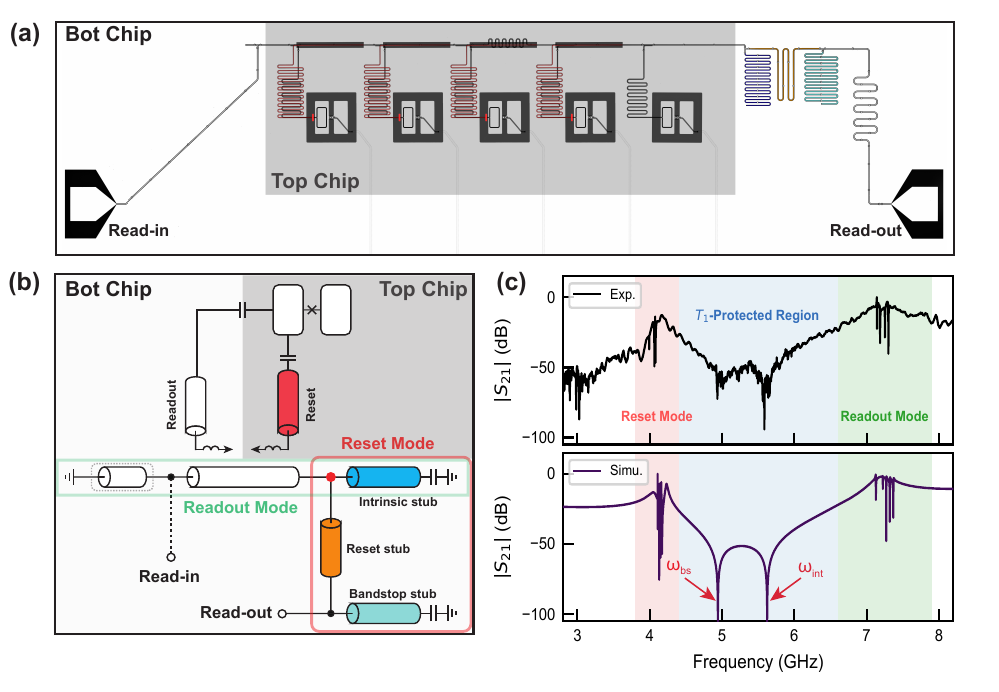}
    \caption{
    \textbf{Filter design.}
    (a) False-color microscope image of the quantum processor, showing the qubit and reset resonators (red) on the top chip (gray background), while the filter and readout resonators reside on the bottom chip.
    (b) Distributed circuit schematic corresponding to (a), detailing the functional components of the architecture. The green border outlines the $3\lambda/4$ coplanar waveguide that defines the readout mode, while the red border outlines the reset mode. The red dot marks the readout node. The read-in port (dashed line) near the grounded end (dashed border) has a negligible response to the filter and resonator parameters and can be omitted in reflection readout, where the probe signal is delivered through the readout port.
    (c) Measured (top) and simulated (bottom) transmission spectra ($S_{21}$) of the filter. The colored shaded regions indicate three frequency bands: a low-frequency reset band near 4.1~GHz, a broadband $T_1$-protected band with two notches at approximately 4.9~GHz and 5.6~GHz, and a high-frequency passband for multiplexed dispersive readout near 7.3~GHz. The readout and reset resonators appear as dips within their respective bands.
    }
    \label{fig:overview}
\end{figure*}

In this work, we experimentally demonstrate a distributed Purcell-filter architecture that incorporates a dedicated fast-reset channel while retaining fast readout and broadband $T_1$ protection.
We combine a bandstop notch and an intrinsic notch into a three-quarter-wavelength ($3\lambda/4$) coplanar waveguide (CPW) bandpass filter for broadband protection, achieving a Purcell-limited $T_{1,\mathrm{P}}$ floor of 1~ms over a 1.2~GHz bandwidth and supporting multiplexed readout with a bandwidth exceeding 300~MHz.
An additional reset stub is inserted between the two notches, creating a reset mode whose frequency is independently tuned by the stub length.
Instead of coupling the qubits directly to a common filter bus—a strategy that risks parasitic crosstalk between qubits—we introduce dedicated reset resonators below the qubit frequencies.
The low-frequency reset resonator enables rapid qubit reset with a decay time constant of approximately 20~ns, while a deliberate $\kappa$--$\chi$ mismatch prevents photon-shot-noise-induced qubit dephasing.
We achieve 100-ns reset with 99.6\% efficiency and suppress photon-shot-noise-induced dephasing by a factor of 70 compared with the readout resonator.
Because the readout and reset bands are independently parameterizable, the filter architecture offers flexible frequency allocation, making it highly practical for scalable superconducting processors.

\section{Filter Design}
\label{sec:device}

\begin{figure}[t]
    \centering
    \includegraphics[width=\columnwidth]{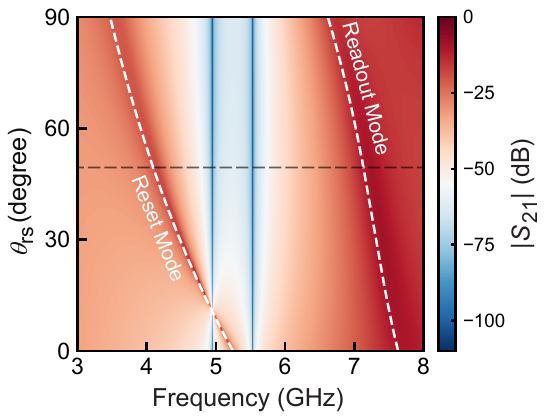}
    \caption{
    \textbf{Simulated transmission response $|S_{21}|$ of the filter versus the electrical length $\theta_{\mathrm{rs}}$ of the reset stub.}
    The electrical length $\theta_{\mathrm{rs}}$ is defined at frequency $\omega_{\mathrm{q}}=5~\mathrm{GHz}$.
    The white dashed lines trace the frequencies of the reset mode (left) and the readout mode (right) as functions of $\theta_{\mathrm{rs}}$.
    The horizontal dashed line marks the design parameter used in this work; the corresponding $S_{21}$ response is shown in Fig.~\ref{fig:overview}(c).
    As $\theta_{\mathrm{rs}}$ is swept, the reset mode frequency shifts significantly more than the readout mode frequency, while the intrinsic and bandstop notch frequencies remain stable.
    }
    \label{fig:S21_vs_TFlen}
\end{figure}

We experimentally implement the Purcell filter in a flip-chip superconducting quantum processor~\cite{foxen2018}, with the experimental setup described in Appendix~\ref{sec:setup}. As shown in Fig.~\ref{fig:overview}(a), the processor consists of a bottom carrier chip hosting the filter structure and readout resonators, and a top chip containing the transmon qubits and reset resonators. The reset resonator is positioned directly above the readout resonator to reduce the on-chip footprint. The corresponding distributed-element circuit model is depicted in Fig.~\ref{fig:overview}(b).

The backbone of the filter is a three-quarter-wavelength ($3\lambda_{\rm r}/4$) coplanar waveguide (outlined by the green border), with a weakly coupled read-in port near the shorted end. The readout node (red dot in Fig.~\ref{fig:overview}(b)), where the backbone connects to the readout port, is positioned at a distance of $\lambda_{\rm int}/4$ from the open end of the backbone, creating a notch at frequency $\omega_{\rm int}$---a feature known as intrinsic filter protection~\cite{sunada2022fast}. A second notch at $\omega_{\rm bs}$ is generated by connecting a $\lambda_{\rm bs}/4$ bandstop stub in parallel to the readout port. The frequencies of the two notches can be independently tuned by adjusting the electrical lengths of the associated stubs. Separating them creates a broadband Purcell-protected band, as confirmed by the measured and simulated $S_{21}$ in Fig.~\ref{fig:overview}(c), where the two transmission zeros at $\omega_{\text{int}}/2\pi = 5.6$~GHz and $\omega_{\text{bs}}/2\pi = 4.9$~GHz correspond to the intrinsic and bandstop notches, respectively.

A reset stub is inserted between the intrinsic and bandstop stubs to form a half-wavelength ($\lambda/2$) reset mode below the qubit frequencies. The resulting filter response comprises three frequency bands: a passband for multiplexed readout above the qubit frequencies, a broadband protection band for the qubits, and a reset band below the qubit frequencies. The extracted linewidths of the readout and reset bands are 300~MHz and 100~MHz, respectively. Adjusting the readout-node position together with the reset-stub length can increase the readout bandwidth to as much as 800 MHz, as discussed in Appendix~\ref{sec:frequency_shift}. A broad readout band is advantageous when uniform resonator linewidths are required for multiplexed readout.

The frequency of the reset mode can be adjusted by the electrical length $\theta_{\mathrm{rs}}$ to satisfy the half-wavelength condition:
\begin{equation}
\theta_{\mathrm{int}}(\omega_{\mathrm{rst}})
+
\theta_{\mathrm{bs}}(\omega_{\mathrm{rst}})
+
\theta_{\mathrm{rs}}(\omega_{\mathrm{rst}})
= 180^\circ.
\label{eq:reset_freq}
\end{equation}
Here, $\theta_{\mathrm{int}}$, $\theta_{\mathrm{bs}}$, and $\theta_{\mathrm{rs}}$ are the electrical lengths of the intrinsic, bandstop, and reset stubs, respectively.
In Fig.~\ref{fig:S21_vs_TFlen}, we show how the frequency of the reset mode changes with the length of the reset stub. Extending $\theta_{\text{rs}}$ introduces a parasitic reactive load on the readout mode, causing a small frequency shift of the readout band.
This shift can be effectively compensated by fine-tuning the electrical length of the grounded section of the filter backbone, see Appendix~\ref{sec:frequency_shift} for details. Such spectral stability ensures that reset-mode tuning does not compromise the integrity of the readout channel.

\begin{figure}[t]
    \centering
    \includegraphics[width=\columnwidth]{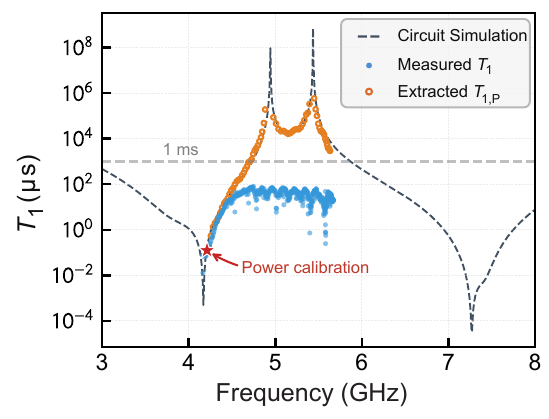}
    \caption{
    \textbf{Experimental characterization of $T_1$ protection.} 
    Measured relaxation time $T_1$ of qubit Q1 (blue dots) and extracted Purcell-limited lifetime $T_{1,\mathrm{P}}$ (orange circles) as a function of frequency.
    Some data points for $T_{1,\mathrm{P}}$ are missing near the notch resonances ($\sim 4.94$~GHz and $\sim 5.46$~GHz) due to insufficient drive strength.
    The black dashed line shows the theoretical prediction from circuit simulation. 
    The calibration reference point is marked by a red star at 4.22~GHz.
    }
    \label{T1-protect}
\end{figure}

\begin{figure*}[t]
    \centering
    \includegraphics[width=1\textwidth]{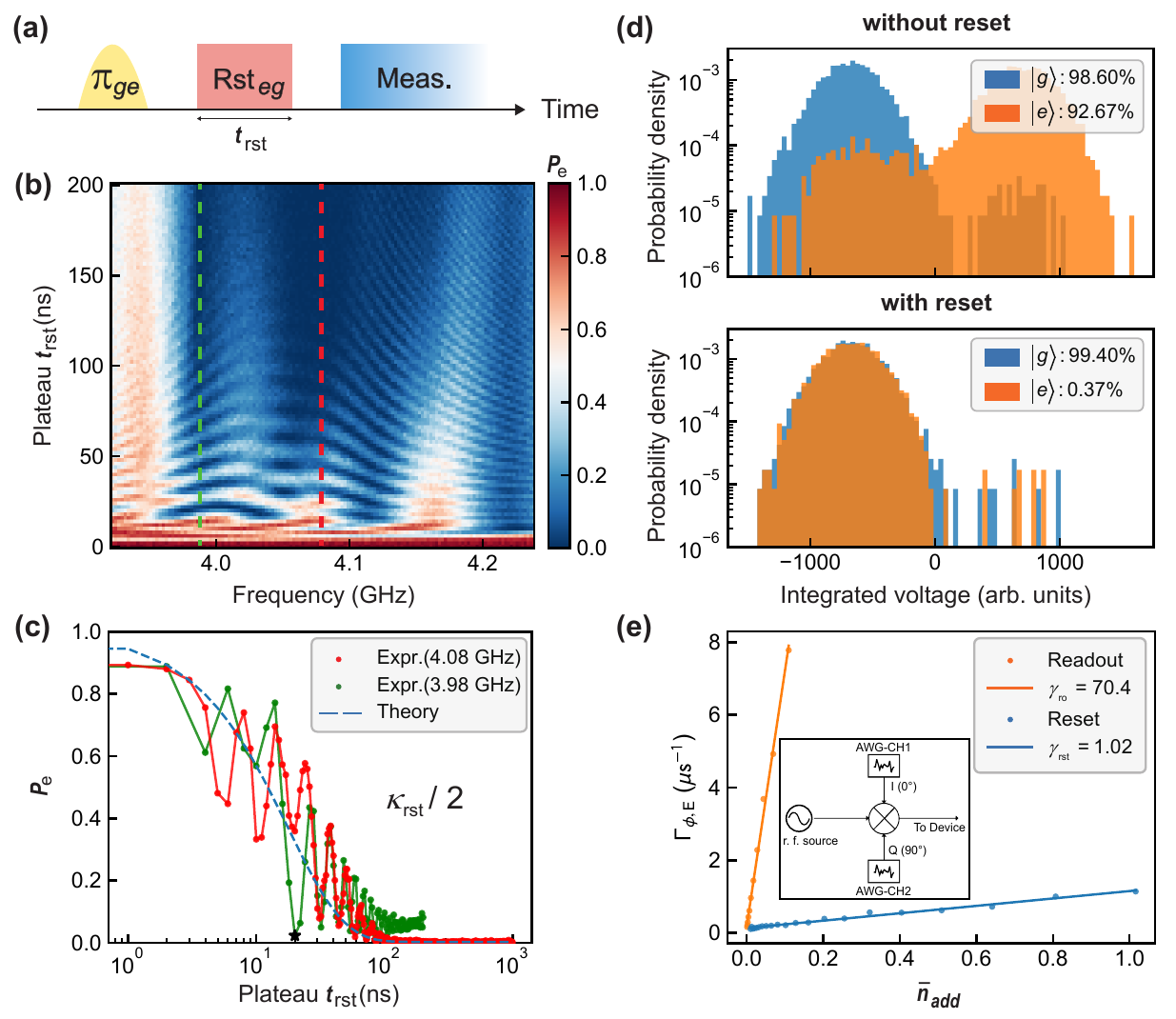}
    \caption{
    \textbf{Reset gate calibration and reset efficiency.} 
    (a) Pulse sequence for calibrating the reset operation. The qubit is prepared in its first excited state and then tuned into resonance with its reset resonator for a variable duration $t_{\mathrm{rst}}$ before measurement.
    (b) Measured chevron pattern for the iSWAP interaction between a qubit and its corresponding reset resonator. The pattern is tilted due to flux pulse distortion. The red dashed line indicates resonance with the reset resonator at 4.07~GHz.
    (c) Line cuts of (b) along the red and green dashed lines. Underdamped oscillations are observed because $g_{q,\mathrm{rst}} > \kappa_{\mathrm{rst}}/4$. The decay envelope of the red curve is dominated by the reset-resonator linewidth, with a decay time constant of approximately 17.6~ns (corresponding to $\kappa_{\rm rst}/2\pi \approx 18$~MHz). The green curve corresponds to a shorter reset operation that exploits the first population minimum.
    (d) Single-shot histograms for the initial $|g\rangle$ (blue) and $|e\rangle$ (orange) states, comparing the cases without (top) and with (bottom) the reset gate. The measurement pulse duration is 500~ns (without Josephson parametric amplifier), and the reset pulse duration is 100~ns. 
    (e) Spin-echo dephasing rate $\Gamma_{\phi,\mathrm{E}}$ at the sweet spot versus added photon number $\bar{n}_{\mathrm{add}}$ in the readout (orange) and reset (blue) resonators. The solid lines are linear fits. These data were acquired on a separate sample with $\kappa_{\mathrm{rst}}/2\pi \approx 14.4$~MHz and $\chi_{\rm rst}/2\pi = 0.16$~MHz. The inset shows the engineered thermal-photon noise source: a coherent tone is mixed with white noise from two independent arbitrary waveform generators (AWGs) applied to the in-phase (I) and quadrature (Q) ports of an I/Q mixer.
    }
    \label{fig:reset}
\end{figure*}

\section{Broadband $T_1$ protection}
\label{sec:III}

The characterized \(T_1\) protection is shown in Fig.~\ref{T1-protect}. The experimentally measured relaxation time \(T_1\) of Q1 in the protected band is approximately \(35~\mu\)s (blue dots in Fig.~\ref{T1-protect}). Unless otherwise noted, all data presented in this paper refer to this example qubit, labeled Q1. However, a direct measurement of \(T_1\) cannot isolate the Purcell-limited lifetime, because the relaxation time contains contributions from multiple decay channels. For instance, the measured \(T_1\) exhibits oscillations arising from standing waves in the drive line, as discussed in Appendix~\ref{app:drive_purcell}.

To isolate the Purcell-limited component, we combine circuit simulation with a Rabi-driving extraction technique~\cite{Nakamura_intrinsic_2022,beaulieu2026fast}. The theoretical Purcell-limited relaxation time is first calculated using the standard expression~\cite{nigg2012black,Matinis_1986,Cleland_2008}
\begin{equation}
    T_{1,\mathrm{P}} = \frac{C_q}{\mathrm{Re}[Y(\omega_q)]},
    \label{eq:Purcell}
\end{equation}
where $C_q$ is the transmon shunt capacitance and $\mathrm{Re}[Y(\omega_q)]$ is the real part of the nodal admittance seen at the qubit port~\cite{solgun2019simple}. The simulated $T_{1,\mathrm{P}}$ is shown as the black dashed line in Fig.~\ref{T1-protect}. Details of the circuit architecture and simulation method are provided in Appendix~\ref{sec:T1loss}.

Experimentally, we apply a resonant microwave drive at $\omega_{\mathrm{q}}$ through the readout port to induce Rabi oscillations. The Purcell-induced decay rate $\Gamma_{\mathrm{p}}$ is then determined from the measured Rabi strength $\Omega$ as
\begin{equation}
\Gamma_{\mathrm{p}} = \frac{\Omega^{2}}{4} \frac{\hbar \omega_{\mathrm{q}}}{P},
\label{eq:PurcellRate}
\end{equation}
where $P$ is the drive power at the qubit transition frequency. To calibrate $P$, we bias the qubit close to the reset mode, where relaxation is dominated by Purcell decay, as indicated by the red star in Fig.~\ref{T1-protect}. The drive power $P$ is then obtained from Eq.~\ref{eq:PurcellRate}. Subsequently, we measure the Rabi strength through the readout port over a frequency range above the reset mode and extract the corresponding $T_{1,\mathrm{P}}$ values.

The extracted $T_{1,\mathrm{P}}$ agrees well with the theoretical prediction. Across a broad bandwidth of approximately 1.2~GHz (4.7--5.9~GHz), the filter maintains a Purcell-limited relaxation time floor of at least 1~ms. This extensive protection window provides substantial operational flexibility for qubit frequency biasing in multi-qubit architectures.

\section{Rapid Reset with dedicated resonator}

Recently proposed multi-purpose architectures couple qubits directly to the filter itself~\cite{ding2025multipurpose,gu2026multimode}, which can introduce filter-mediated stray interactions between qubits. To avoid this, we introduce a dedicated reset resonator for each qubit and design it such that $g_{q,\rm rst} > \kappa_{\rm rst}/4$ and $\kappa_{\rm rst} \gg 2\chi_{\rm rst}$. This design ensures rapid reset, suppresses unwanted qubit interactions during idle operation, and provides robustness against photon-shot noise. For our device, the reset resonator linewidth is $\kappa_{\text{rst}}/2\pi \approx 18$~MHz and the qubit-resonator coupling strength is $g_{\text{q,rst}}/2\pi \approx 32$~MHz.

Qubit reset is realized by tuning the qubit into resonance with its reset resonator using a square pulse of duration $t_{\rm rst}$. The chevron pattern in Fig.~\ref{fig:reset}(b) shows the resulting iSWAP interaction. Because $g_{q,\rm rst} > \kappa_{\rm rst}/4$, the system operates in the underdamped regime. The reset speed is limited by the reset-resonator linewidth; with $\kappa_{\text{rst}}/2\pi \approx 18$~MHz, the qubit excitation decays to the baseline within 100~ns (Fig.~\ref{fig:reset}(c)). Alternatively, the reset pulse can be shortened to approximately 20~ns by ending it at the first minimum of the qubit excited-state population during the excitation swap (green curve in Fig.~\ref{fig:reset}(c)).

We quantify the reset gate efficiency using a single-shot readout experiment. 
With a reset pulse of $t_{\rm rst} = 100$~ns (more than ten times the reset-resonator decay time), the excited-state population is reduced from $P_e = 92.67\%$ to $P_e^{\mathrm{rst}} = 0.37\%$, corresponding to a reset efficiency of $\eta_{\mathrm{rst}} = 1 - P_e^{\mathrm{rst}}/P_e = 99.6\%$ (Fig.~\ref{fig:reset}(d)). The residual population is primarily due to readout-induced state transitions and the finite thermal occupation of the reset resonator. 
Low-frequency reset resonators ($\sim 4$~GHz) tend to have higher thermal occupations according to the Boltzmann distribution. To mitigate this, we add 5~GHz high-pass filters (HPFs) in the read-in and readout lines (see Appendix~~\ref{app:high-pass-filter}), which prevent thermal photons from entering the reset resonator. Without these filters, the residual thermal population rises to $1.87\%$ in a separate cool-down (Appendix~\ref{app:high-pass-filter}).

Even with finite thermal occupation, the intentional $\kappa$--$\chi$ mismatch ensures that residual photons do not induce significant dephasing. We quantify this using the noise-injection protocol of Ref.~\cite{yan_flux_2016}. White noise is mixed with a coherent tone and injected into the resonator, creating a controllable thermal-photon population $\bar{n}_{\mathrm{add}}$. The resulting pure dephasing rate $\Gamma_{\phi,\mathrm{E}}$, measured at the qubit sweet spot, scales linearly with $\bar{n}_{\mathrm{add}}$ for both readout and reset resonators (Fig.~\ref{fig:reset}(e)). From the slopes, the reset resonator exhibits a dephasing sensitivity of $\gamma_{\mathrm{rst}} \approx 1.02~\mu\mathrm{s}^{-1}$, about 70 times lower than that of the readout resonator ($\gamma_{\mathrm{ro}} \approx 70.4~\mu\mathrm{s}^{-1}$). 
This substantial suppression of photon back-action confirms that the architecture enables fast reset without compromising qubit coherence. The reduced dephasing penalty from residual reset photons may allow subsequent operations to begin before the reset resonator is fully depleted, particularly when excitation swapping is exploited for 10-ns-scale reset.

\section{Conclusion}
\label{sec:conclusion}

We have demonstrated a Purcell filter that integrates broadband $T_1$ protection, fast readout, and fast reset within a simple distributed microwave structure. The readout, protection, and reset bands are independently parameterizable, offering greater design flexibility than previous multi-mode filters~\cite{gu2026multimode}.
The filter provides a Purcell-limited relaxation time exceeding $1$~ms over a $1.2$~GHz bandwidth, enabling single-shot readout with a measurement time of $500$~ns, without using a Josephson parametric amplifier.
In contrast to previous multi-purpose architectures that rely on direct qubit-filter coupling and suffer from stray interactions~\cite{ding2025multipurpose,gu2026multimode}, we employ dedicated reset resonators for rapid reset. These low-frequency reset resonators achieve 100~ns reset with $99.6\%$ efficiency. An intentional $\kappa$--$\chi$ mismatch suppresses photon-shot-noise-induced dephasing by a factor of approximately 70 relative to the standard readout path, eliminating the need to wait for resonator photon depletion and opening a practical route to 10-ns-scale reset.
These features make the architecture highly practical for scalable superconducting processors.

\begin{acknowledgments}
This work was supported by the National Natural Science Foundation of China (123b2071), and the Innovation Program for Quantum Science and Technology (2021ZD0301703).
\end{acknowledgments}

\bibliography{refs}
\bibliographystyle{apsrev4-2}.

\appendix

\begin{figure*}[t] 
    \centering
    \includegraphics[width=0.95\textwidth]{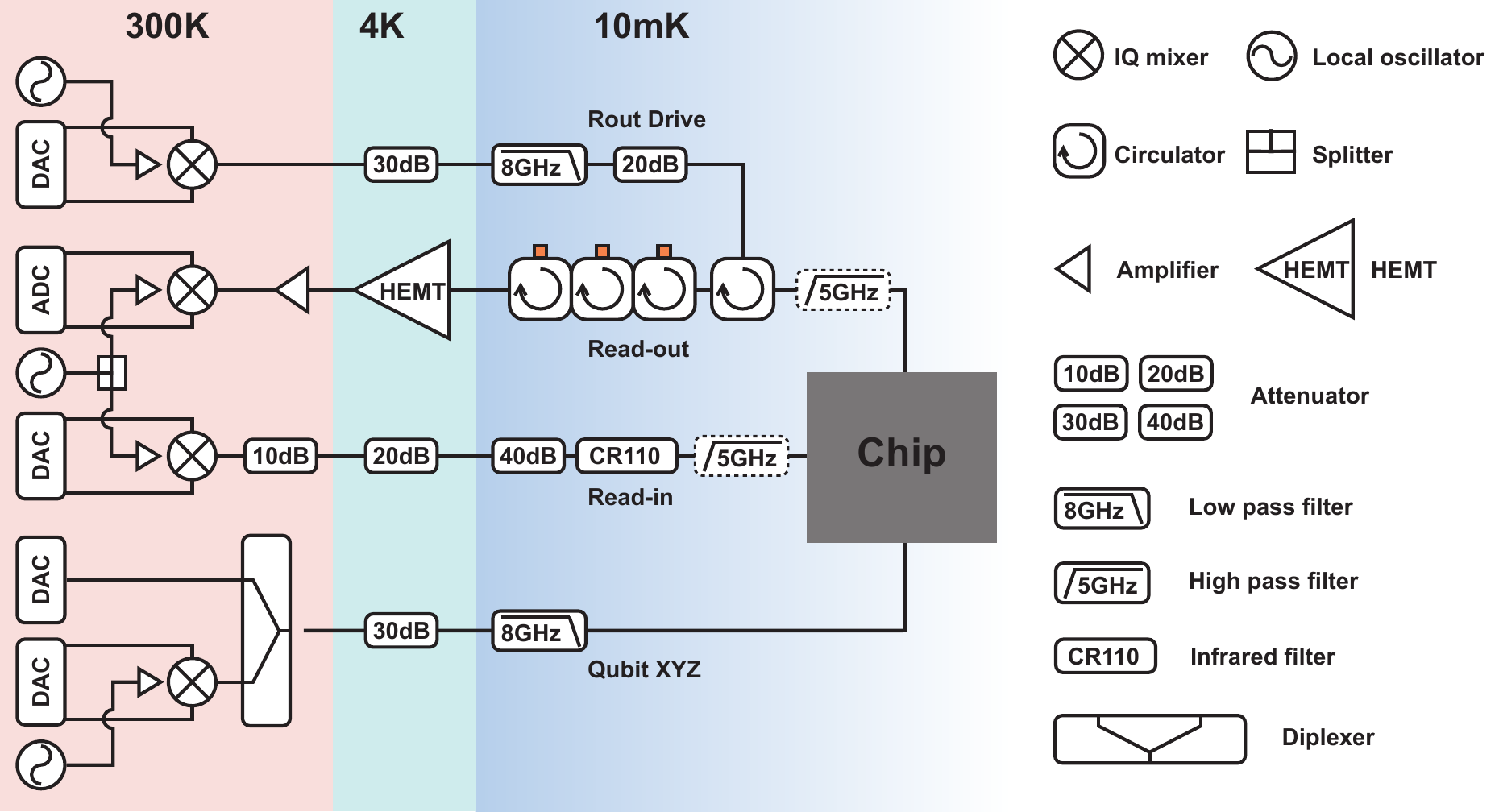} 
    \caption{
    Experimental setup. An additional drive line connected to the circulator in the readout chain is used to characterize the Purcell protection of the device. In other characterization experiments, drive signals are delivered to the qubits through dedicated control lines. To suppress the thermal population of the reset resonator, 5~GHz HPFs (indicated by dashed borders) are added to the read-in and readout lines during reset efficiency characterization.
    }
    \label{fig:experimental_setup}
\end{figure*}

\begin{table}[htbp]
\centering
\renewcommand{\arraystretch}{1.15}
\begin{tabular*}{\columnwidth}{@{\extracolsep{\fill}}lcc@{}}
\toprule[1.2pt]
Parameters & Q1 \\ 
\midrule
$\omega_{\mathrm{q},\mathrm{max}}/2\pi$~(GHz)     & $5.664$   \\
$\alpha/2\pi$~(MHz)                      & $-210$ \\
$\omega_{\mathrm{r}}/2\pi$~(GHz)   & $7.28$ \\
$\kappa_{\mathrm{r}}/2\pi$~(MHz)      & $7.58$  \\
$\omega_{\mathrm{rst}}/2\pi$~(GHz)        & $4.08$  \\
$\kappa_{\mathrm{rst}}/2\pi$~(MHz)        & $18.0$   \\
$T_{1,\mathrm{median}}$~($\mu$s)                 & $34.2$ \\
\bottomrule[1.2pt]
\end{tabular*}
\caption{Parameters of qubit $Q_1$ used in the experiments, together with the corresponding readout and reset parameters. The median $T_1$ is measured over the range 4.4--5.6~GHz (protection region).}
\label{tab:system_params}
\end{table}

\section{Experiment Setup and Device Parameters}
\label{sec:setup}

Our experiment is performed on a flip-chip superconducting processor consisting of a bottom sapphire chip coated with aluminum and a top sapphire chip coated with tantalum.
The flux-tunable transmon qubits and the reset resonators are fabricated on the top chip, while the filter and the readout resonators are on the bottom chip. The device is cooled to 10~mK in a dilution refrigerator and measured using the setup shown in Fig.~\ref{fig:experimental_setup}.
To characterize the Purcell protection of the device, an additional drive line is connected to the circulator in the readout chain, through which the Rabi drive signal is delivered to the qubits. In other characterization experiments, drive signals are sent to the qubits through dedicated control lines. The data presented in Fig.~\ref{T1-protect} and Figs.~\ref{fig:reset}(a)--(d) are obtained from Q1, with the corresponding parameters summarized in Table~\ref{tab:system_params}.

\begin{figure}[htbp]  
    \centering
    \includegraphics[width=\columnwidth]{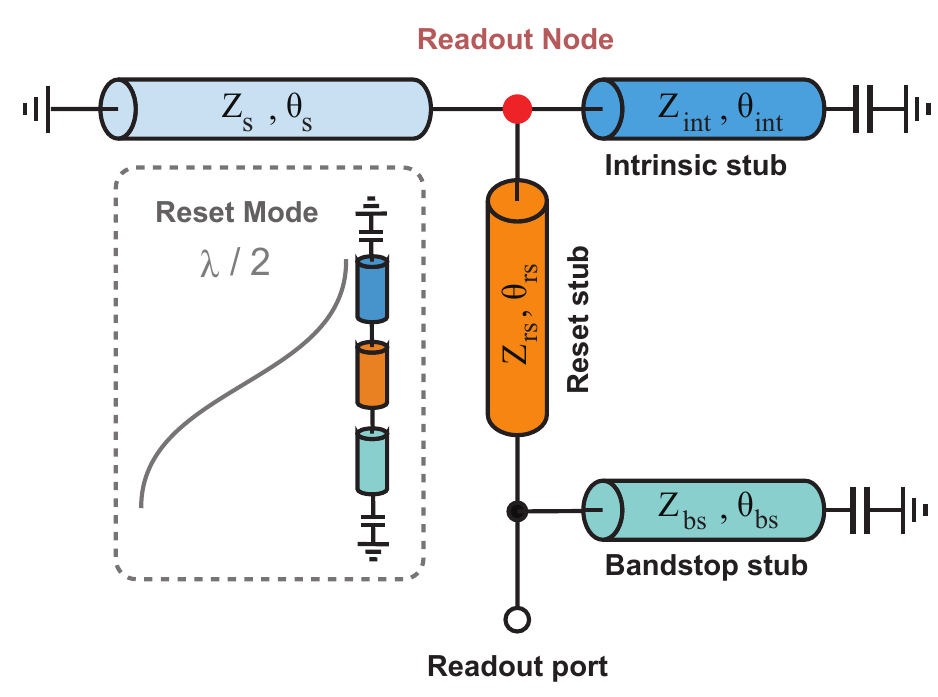}
    \caption{
    Circuit model of the integrated Purcell filter. The reset mode is formed by a half-wavelength ($\lambda_{\rm r}/2$) coplanar waveguide with two open ends, comprising the intrinsic stub, the bandstop stub, and the reset stub in between. The red dot marks the readout node.
    }
    \label{fig:appendix_circuit}  
\end{figure}

\section{Frequency and bandwidth of the Reset and Readout Mode}
\label{sec:frequency_shift}

\subsection{Reset-Mode Frequency}
\label{sec:reset_shift}

The reset mode can be understood from the accumulated electrical phase along the reset path, which is formed by the intrinsic stub, the reset stub, and the bandstop stub. We denote their electrical lengths by
\[
\theta_{\mathrm {int}}(\omega),\qquad
\theta_{\mathrm{rs}}(\omega),\qquad
\theta_{\mathrm{bs}}(\omega),
\]
respectively. The reset mode corresponds to a half-wavelength ($\lambda_{\rm r}/2$) coplanar waveguide with two open ends, so the total accumulated electrical length approximately satisfies
\begin{equation}
\Theta_{\mathrm{rst}}(\omega_{\mathrm{rst}})
=
\theta_{\mathrm {int}}(\omega_{\mathrm{rst}})
+
\theta_{\mathrm{rs}}(\omega_{\mathrm{rst}})
+
\theta_{\mathrm{bs}}(\omega_{\mathrm{rst}}) =180^\circ.
\label{eq:Sm_Theta_rs}
\end{equation}
Here $\omega_{\mathrm{rst}}$ is the reset-mode frequency. Since the lengths of the intrinsic and bandstop stubs are fixed by the $T_1$ protection requirement, the reset mode frequency can be tuned by adjusting the length of the reset stub.

\begin{figure}[htbp] 
    \centering
    \includegraphics[width=\columnwidth]{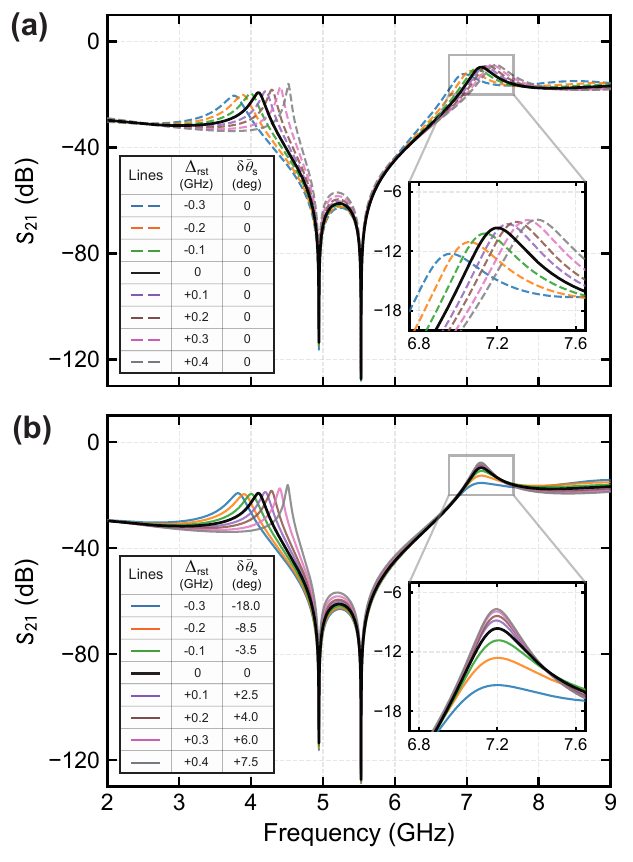} 
    \caption{
    \textbf{Compensation of the readout-mode frequency when tuning the reset-mode frequency.} (a) Filter response for various reset-mode frequencies without compensation of the coupling-section electrical length. (b) Filter response with compensation applied ($\delta\bar{\theta}_{\mathrm{s}}$), showing improved stability of the readout mode.
    }
    \label{fig:appendix_comp}
\end{figure}

The bandwidth of the reset mode is determined by the electrical length $\theta_{\rm bs}$ of the bandstop stub. As the reset mode frequency approaches the bandstop notch frequency, the readout port position moves closer to a voltage node, leading to a higher quality factor for the reset mode, as shown in Fig.~\ref{fig:appendix_comp}.

\subsection{Readout-Mode Frequency}
\label{sec:load_effect}

The bare readout resonance is defined by the filter backbone alone. At the readout node (red point in Fig.~\ref{fig:appendix_circuit}), the backbone is modeled as the parallel combination of a short-circuited section \((Z_{\mathrm s},\theta_{\mathrm s})\) and an open-circuited section \((Z_{\mathrm {int}},\theta_{\mathrm {int}})\). The bare nodal admittance is
\begin{equation}
    Y_{\mathrm r}(\omega)
    =
    -\frac{j}{Z_{\mathrm s}}\cot\theta_{\mathrm s}(\omega)
    +
    \frac{j}{Z_{\mathrm int}}\tan\theta_{\mathrm {int}}(\omega),
    \label{eq:S_readout_bare}
\end{equation}
and the bare readout frequency \(\omega_{\mathrm r}\) is determined by
\begin{equation}
    \Im\!\left[
    Y_{\mathrm r}(\omega_{\mathrm r})
    \right]
    =0.
\end{equation}

When the reset branch is attached, it contributes a transformed load impedance \(Z_{\mathrm L}\) to the same node. The reset branch consists of the reset section \((Z_{\mathrm{rs}},\theta_{\mathrm{rs}})\) terminated by an open-circuited bandstop stub \((Z_{\mathrm{bs}},\theta_{\mathrm{bs}})\). The load impedance seen from the readout node is
\begin{equation}
    Z_{\mathrm L}(\omega)= Z_{\mathrm{rs}}
    \frac{Z_{\mathrm{bs}}^{\mathrm{in}}(\omega) + j Z_{\mathrm{rs}}\tan\theta_{\mathrm{rs}}(\omega)
    }{Z_{\mathrm{rs}} + j Z_{\mathrm{bs}}^{\mathrm{in}}(\omega)\tan\theta_{\mathrm{rs}}(\omega)},
    \label{eq:S_load_impedance}
\end{equation}
where
\begin{equation}
    Z_{\mathrm{bs}}^{\mathrm{in}}(\omega)
    = -j Z_{\mathrm{bs}}\cot\theta_{\mathrm{bs}}(\omega).
\end{equation}
The loaded readout resonance is then determined by
\begin{equation}
    \Im\!\left[
    Y_{\mathrm r}(\omega)+Y_{\mathrm L}(\omega)
    \right]
    =0,
\end{equation}
with $Y_{\mathrm L}(\omega)  = 1/Z_{\mathrm L}(\omega)$.

\subsection{Compensation of Readout-mode frequency}
\label{sec:compensation_protocol}

We introduce a compensation protocol that restores the readout mode to its target frequency when the reset mode frequency is adjusted. The compensation is implemented by adjusting the electrical length \(\theta_s\) of the grounded side of the filter backbone, with the correction denoted by \(\delta{\theta}_{\mathrm {s}}\). The compensation condition is obtained by requiring that the total imaginary nodal admittance remain zero at the unperturbed readout frequency \(\omega_{\mathrm r}\):
\begin{equation}
    \Im\!\left[
    Y_{\mathrm r}(\theta_{\mathrm s}+\delta{\theta}_{\mathrm {s}},\omega_{\mathrm r})
    +Y_{\mathrm L}(\omega_{\mathrm r} ) + \delta Y_{\mathrm L}(\omega_{\mathrm r} )
    \right]
    =0.
    \label{eq:S_admittance_compensation}
\end{equation}
Here \(\delta Y_{\mathrm L}(\omega_{\mathrm r} )\) is the change in load admittance when the reset mode frequency is adjusted.

Using Eqs.~\eqref{eq:S_readout_bare} and \eqref{eq:S_load_impedance}, the required compensation length is obtained as
\begin{equation}
    \delta {\theta}_{\mathrm s}
    \approx
    -\,Z_{\mathrm s}\sin^{2}( \theta_{\mathrm s})\,
    \Im\!\left[
    \delta Y_{\mathrm L}(\omega_{\mathrm r})
    \right].
    \label{eq:S_compensation_length}
\end{equation}
Equation~\eqref{eq:S_compensation_length} provides a direct mapping from the residual reactive load of the reset branch to the short-side correction required to cancel it.

Figure~\ref{fig:appendix_comp}(b) shows that, with compensation, the readout-mode frequency remains unchanged as the reset-mode frequency is varied from 3.8 to 4.5~GHz. Since the compensation section changes the electrical position of the readout node, a positive (negative) compensation length reduces (increases) the readout-mode linewidth, as shown in the inset of Fig.~\ref{fig:appendix_comp}(b). A large readout-mode linewidth, as indicated by the blue line, is advantageous when uniform resonator linewidths are required for multiplexed readout.

\section{Drive-line induced loss}
\label{app:drive_purcell}

\begin{figure}[htbp]
    \centering
    \includegraphics[width=\columnwidth]{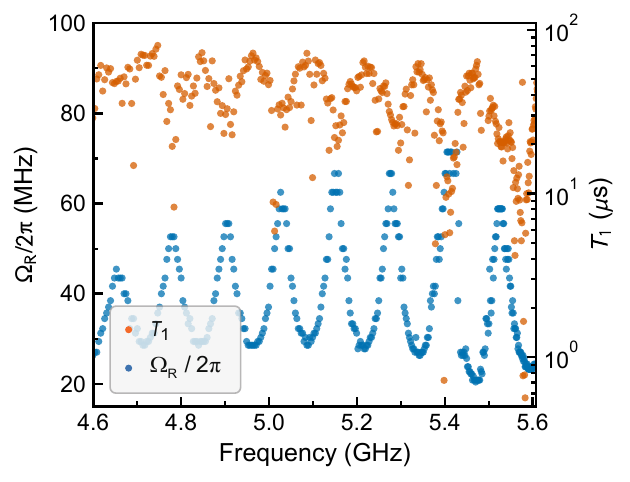}
    \caption{Measured $T_1$ and Rabi drive strength of Q1 versus qubit frequency. The oscillation period is 110~MHz, consistent with a 90-cm-long coaxial drive line anchored at the mixing chamber stage.}
    \label{fig:rabi_drive}
\end{figure}

We observe pronounced frequency-dependent oscillations in both the qubit relaxation time ($T_1$) and the Rabi drive strength ($\Omega_{\text{R}}/2\pi$), as shown in Fig.~\ref{fig:rabi_drive}. Across the measured frequency band, the two quantities are clearly anti-correlated: a stronger drive corresponds to a shorter $T_1$~\cite{Houck2008_PRL}. The oscillatory features exhibit a common modulation period of approximately $110$~MHz for both $\Omega_{\text{R}}/2\pi$ and $T_1$. Notably, this periodicity coincides with the free spectral range of a standing wave along the 90-cm-long coaxial drive line anchored at the mixing chamber (MXC) stage, indicating that the standing wave arises from impedance mismatch of the infrared filter at the MXC. These results indicate that the intrinsic $T_1$ of this sample is limited by the drive-line-induced loss.

\section{SPICE Simulation of the Purcell-Limited $T_1$}
\label{sec:T1loss}

\begin{figure}[htbp]
    \centering
    \includegraphics[width=0.48\textwidth]{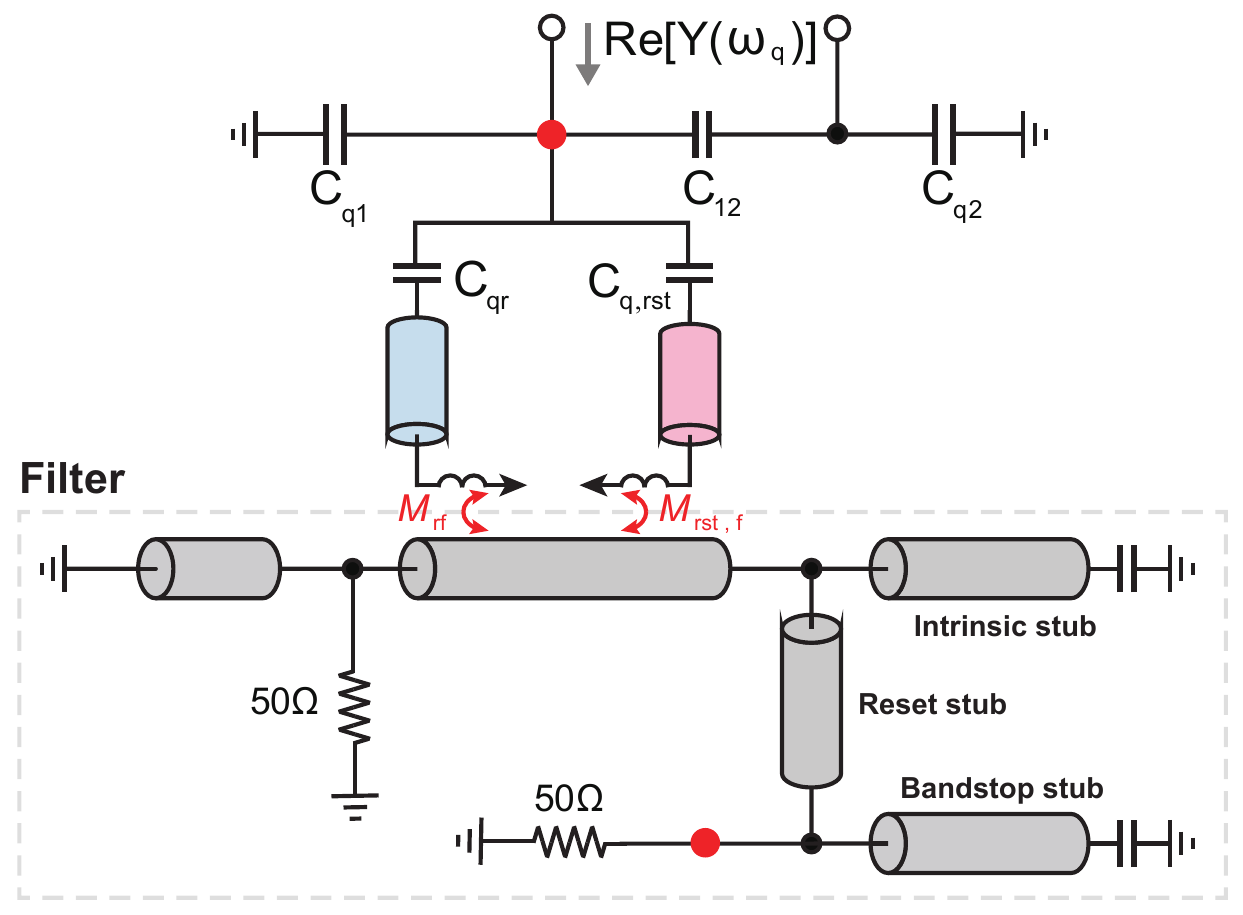}
    \caption{Circuit model for the SPICE simulation of the Purcell-limited $T_1$. The readout and read-in ports are terminated by $50~\Omega$ loads.}
    \label{fig:T1loss}
\end{figure}

The Purcell-limited relaxation time of the qubit shown in Fig.~\ref{T1-protect} is calculated using SPICE simulation, with the circuit model depicted in Fig.~\ref{fig:T1loss}. The qubit relaxation rate to the environment is determined by the real part of the admittance seen at the qubit port~\cite{nigg2012black}, yielding
\begin{equation}
    T_{1,\mathrm{P}}(\omega_q)
    =
    \frac{C_{\Sigma}}
         {\mathrm{Re}\left[Y(\omega_q)\right]}.
    \label{eq:t1_admittance}
\end{equation}
This simulation captures the radiative loss channels through the readout chain, including losses mediated by both the readout and reset resonators.

\section{Improved Reset Efficiency with HPFs}
\label{app:high-pass-filter}

\begin{figure}[htbp]
    \centering
    \includegraphics[width=\columnwidth]{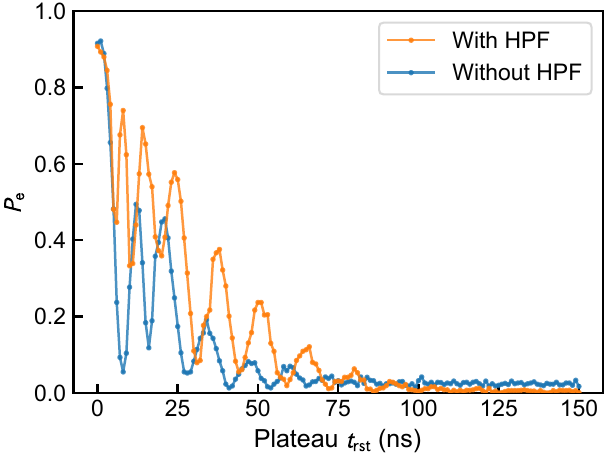}
    \caption{
    Measured decay envelope during the reset operation. The orange (blue) curve corresponds to data with (without) 5~GHz HPFs added to the read-in and readout lines, showing a residual population of 0.37\% (1.87\%).
    }
    \label{fig:rabi_drive}
\end{figure}

We measure the reset efficiency of the same qubit across different cool-downs. With 5~GHz HPFs added to the read-in and readout lines, we observe a much lower residual population after 100~ns, indicating a reduced thermal population of the reset resonator.

\end{document}